\documentstyle[aps,preprint,psfig]{revtex}
\begin{document}
\draft
\preprint{Abst. No. 568, Prog. Nr. SS-ThM5}
\title{Mechanism of efficient carbon monoxide oxidation at Ru\,(0001)}
\author{C. Stampfl and M. Scheffler}
\address{
Fritz-Haber-Institut der Max-Planck-Gesellschaft,
Faradayweg 4-6, D-14 195 Berlin-Dahlem, Germany
}
\date{\today}


\maketitle

\vspace*{-10pt}
\vspace*{-0.7cm}
\begin{quote}
\parbox{16cm}{\small
We performed density-functional theory calculations using the
generalized gradient approximation for the exhange-correlation functional to 
investigate the unusual catalytic behavior of Ru under
elevated gas pressure conditions for the
carbon monoxide oxidation reaction, which includes a particularly
high CO$_{2}$ turnover.
Our calculations indicate that a full monolayer of adsorbed
oxygen actuates the high
rate, enabling CO$_{2}$ formation via both scattering of
gas-phase CO molecules as well as by CO molecules adsorbed
at oxygen vacancies in the adlayer, where the latter mechanism
is expected to be very efficient due to the relatively weak adsorption energy
of {\em both} CO and O, as well as the close proximity of these reactants.
In the present paper we analyse
the bonding and electronic properties
associated with the reaction pathway for CO$_{2}$ production
via the scattering reaction. We find that the identified ``bent'' transition
state is due to electron transfer into the unoccupied 2$\pi$ orbitals
of the CO molecule which reduces the Pauli repulsion between the impinging
CO and the O-covered surface. Bond formation to CO$_{2}$ then proceeds by electron 
transfer back from the CO 2$\pi$ orbitals into the bonding region 
between CO and the adsorbed O atom.
} 
\end{quote}
 
\newpage
\section{Introduction}
Due to its fundamental importance and relative simplicity,
the oxidation of carbon monoxide
may be regarded as a prototype or model system
of a heterogeneous catalytic reaction
(molecular and dissociative (atomic) adsorption, surface reaction,
desorption of products). This reaction has 
thus been extensively studied, and in many regards a good understanding
has been established (see, for example
Refs.~\cite{engel,king,peden1} and references therein). To date, the reaction
has always been found to proceed via 
the Langmuir-Hinshelwood (L-H) mechanism, i.e.,
reaction occurs between constituents that are both chemisorbed on, and in
thermal equilibrium with, the surface.
However, particularly interesting experimental results from recent elevated gas 
pressure catalytic reactor studies
(e.g. operating  at $\sim$ 10 torr with CO/O$_{2}$ pressure ratios $ < 1$)
have been reported for the CO oxidation reaction
over Ru\,(0001) \cite{peden1,peden3} which
led to the proposal that the Eley-Rideal (E-R) mechanism may be operational.
In the E-R mechanism, the reaction takes place between
 particles in the {\em gas-phase} and particles chemisorbed on
the surface.
So far, E-R mechanisms, although discussed and speculated upon
for the CO oxidation reaction in the past, have never been
experimentally confirmed for this reaction. 
Firstly, these recent results found that the rate of CO$_{2}$ production
is significantly higher than at other
transition metal surfaces; in contrast, under ultra high 
vacuum (UHV) conditions, Ru\,(0001) is notably
the poorest catalyst for this reaction~\cite{king}.
Secondly, the measured kinetic data 
is markedly different to that of
other substrates, and as opposed to other transition metal 
catalysts (Rh, Pt, Pd, and  Ir), 
 highest rates of CO$_{2}$ formation
 occur for {\em high} surface oxygen concentrations. 
Finally, almost {\em no} chemisorbed CO was detected during or after
the reaction.  
The experimental findings described above for CO oxidation at
Ru\,(0001)  place a question mark
over the reaction for this system; 
-- in addition, one can also be placed over 
a {\em microscopic} understanding of this
most basic heterogeneous catalytic reaction because to date it does not exist. This is
largely because experiments are not readily
able to probe the reaction from start to finish, i.e. they
cannot follow the reaction pathway in microscopic detail; 
rather they offer information prior, and subsequent, to the reaction.
In this respect theory can offer a way to obtain significant insight
into how the reaction proceeds.

The majority of theoretical investigations of
reactions  at surfaces performed to date have used
a cluster geometry, see e.g. Refs.~\cite{baerends,whitten} and
references therein. In order to reduce (or avoid) possible problems with 
artificially localized wave functions and with cluster boundary
conditions, 
we choose to use an extended surface by
employing the supercell approach.  The cell contains
a number of atomic layers and a region of vacuum space.
Thus the surface is extended in the $x$- and $y$-directions
and is periodic on the scale of the supercell. The periodicity
perpendicular to the surface
should not play a role if the vacuum region is taken large
enough. Such a supercell may thus be called a cluster where, however,
in contrast to standard cluster calculations, the boundary
conditions are treated chemically appropriately and the size of our
``cluster'' is significantly larger. In fact, due to the proper
treatement of the boundary conditions, the results converge very
fast and {\em steadily} with the cell size.

Numerous calculations have demonstrated
that density functional theory (DFT) within the local density approximation
(LDA) represents a good description of the
quantum-mechanical many-body interactions
for well-bonded situations.
For less well-bonded situations, for example, transition-state geometries
where bonds are in the process of being broken
and reformed, e.g. dissociative adsorption and associative desorption, 
or for systems containing strong charge
inhomogenuities, e.g. molecules at
surfaces, recent results have revealed that
the LDA can lead to inaccuracies.
In this respect, using a generalized gradient approximation
(GGA) for the exchange-correlation functional has been shown to
significantly improve upon results obtained using the LDA
(see for example,
Refs.~\cite{philipsen,hu1,hammer-al110-2,pederson}).
Some questions, however, have recently been raised concerning the accuracy
of the GGA for the potential energy surfaces (PES's) for H$_{2}$ dissociation
\cite{nachtigall}. Nevertheless, the GGA is the best justified treatment for the
exchange-correlation functional to date  and we therefore
choose to employ it in our present study.
We like to note that due
 to recent theoretical developments and improved calculation methods
as well as faster computers, it is only in the last couple
of years that calculations
of the type reported here have become possible 
for reactions at surfaces;
the studies perfomed to date
have so far focused only on the ``simple'' dissociative adsorption
of H$_{2}$.
The present study, the main results of which
were reported earlier \cite{stampfl-co}, represents
the first investigation of a
``real'' {\em surface chemical reaction} (to be distinguished from dissociative
chemisorption) using such theoretical methods.

In this paper we briefly describe the main results of our study and discuss
the bonding and electronic structure of the identified reaction
pathway for CO$_{2}$ formation via scattering of CO from the gas phase
$(g)$ and adsorbed $(a)$ oxygen, i.e.,
CO$(g)$ + $\frac{1}{2}$O$_{2}(a) \longrightarrow$ CO$_{2}(g)$.

\section{Calculation Method}
We use density-functional theory (DFT)
with the pseudopotential plane wave method and the supercell approach
\cite{neugebauer,stumpf}.
The surface is modelled using a $(2 \times 2)$ surface unit cell
with four layers of Ru\,(0001) and a vacuum region equivalent
to thirteen such layers.
The {\em ab initio} fully separable pseudopotentials were
created using the scheme of Troullier and
Martins~\cite{troullier}. 
For the calculations discussed in the present paper, the
plane wave cut-off was taken to be 40 Ry, 
with three {\bf k}-points \cite{cunningham} in the surface Brillouin zone.
The GGA of Perdew 
{\em et al.}~\cite{perdew} was employed
for the exchange-correlation functional and it was used in creating
the pseudopotential
as well as in the total-energy functional. It is therefore
treated in a consistent way, from the free atom to the
solid surface and the reactants.
The adsorbate structures were  created on one side of the slab~\cite{neugebauer}
where
the position of all atoms were allowed to relax, except the bottom two Ru 
layers which were held at their bulk positions.
For further details we refer to Refs.~\cite{stampfl-co,stampfl1,stampfl2,over}.

\section{CO at $(1 \times 1)$-O/Ru\,(0001)}
Earlier we reported results of
DFT-GGA calculations for various adsorbate
phases of oxygen on Ru\,(0001)~\cite{stampfl1}
and found that a structure with a ($1 \times 1$) periodicity
and coverage  $\Theta = 1$ should be stable on the surface where
the oxygen atoms occupy the hcp hollow sites.
(Here the coverage $\Theta$
is defined as the ratio of the
concentration of adparticles to that of substrate atoms in the topmost
layer.)
This result indicated that
the formation of the $\Theta = 1$ structure under UHV conditions
from gas phase O$_2$ is kinetically hindered,
but that by using {\em atomic} oxygen (or high O$_{2}$ gas pressures),
this phase should be attainable.
Indeed this structure was successfully created under UHV
by starting from the $(2 \times 1)$ phase with coverage $\Theta=1/2$, 
obtained by exposing Ru\,(0001) to O$_{2}$ at room temperature,
and then offering additional
atomic oxygen to the surface via the dissociation of NO$_2$~\cite{over}.
The determined surface atomic structure of the $(1 \times 1)$ phase by
the low-energy electron diffraction (LEED) intensity analysis
of Ref.~\cite{over}
agreed very well with that predicted by the DFT-GGA calculations.
Furthermore, the calculations showed that the adsorption
energy of oxygen in this phase is significantly weaker than in the
lower coverage ordered phases of $(2 \times 1)$ and $(2 \times 2)$
periodicities; compare $-$1.89~eV to $-$2.33~eV and $-$2.60~eV, respectively,
where the reference energy is taken to be that of $\frac{1}{2}$O$_{2}$.
We note that
evidence for higher oxygen coverages on Ru\,(0001) had been indicated
by earlier experimental 
work; in particular, it had been proposed by Malik and Hrbek \cite{malik}
that a {\em three}  monolayer
structure may form.
Our more recent work \cite{over}, and that of Mitchell and
Weinberg \cite{weinberg}, however, showed that this structure is not
realised. Instead, using NO$_{2}$, additional oxygen enters the 
bulk region after
completion of the $(1 \times 1)$ O adlayer.
Also, in the CO oxidation experiments performed at
elevated gas pressures as described in the introduction,
auger electron spectroscopy (AES) measurements
indicated that there was one monolayer at
the surface. 

In our theoretical study, we therefore initially 
assumed that the  $(1 \times 1)$ phase covered the surface
and investigated
the interaction of CO with this surface. To do this we
calculated the energy for CO at various lateral positions 
and vertical positions over $(1 \times 1)$-O/Ru\,(0001), i.e., we mapped
the total energy on a three-dimensional grid over the surface.
The results showed that CO {\em cannot} 
adsorb on the surface,
-- thus the L-H mechanism is completely inhibited by the presence of a
 perfectly ordered $(1 \times 1)$ O  adlayer.
The next step of our study was the 
consideration of an E-R mechanism as had been speculated 
upon on the basis of the experimental results. 
In order to do this, 
we evaluated an appropriate cut through the high-dimensional
potential energy surface (PES); 
this cut is defined by the
vertical position of the C atom 
and the vertical position of the O adatom below the molecule.
The resulting PES is presented in Fig.~1.
In these calculations,
the CO axis was held perpendicular to the surface
(i.e., a ``collinear'' geometry) in order to ease the analysis. 
This restricted geometry was later
released.  At each point in the PES, the positions
of all the atoms were relaxed except the
bottom two Ru layers and the position of the C and O adatom which
are fixed. From our results 
a reaction pathway for CO$_{2}$ formation via the E-R mechanism could
be identified, where
the associated activation energy barrier is about 1.6~eV.
The PES in Fig.~1,
 indicates that CO$_{2}$ formation is achieved via an upward movement
of the O adatom by $\approx 0.4$ \AA\, towards the incoming CO molecule 
when the molecule is in the range of 2.1~\AA \,to 2.4~\AA\, away from
the surface.
Because of the similar masses of C and O 
it is quite conceivable that a relatively energetic
impinging CO molecule will impart a significant amount of energy to the O
adatom, thus stimulating its vibrations and facilitating its motion
(as indicated by the oscillations in the dot-dashed curve of Fig.~1).
[In this respect, we like to point out that the present system is notably 
different to H$_{2}$ dissociation in which  the significant
mass missmatch between the H atoms and atoms of the substrate
makes it is reasonable to keep the substrate atoms rigid during
the dissociation process. Due to the heavier
reactant species of the present study, 
it is important to consider substrate
relaxation.] Such movement
would then  bring the system to the transition state of the reaction
as denoted by the asterisk in Fig.~1.
The calculation shows that the newly formed CO$_{2}$ molecule is
strongly repelled
from the surface and moves into the vacuum region 
with a significant energy gain of 1.95~eV.

On releasing  the constraint 
that the CO axis  be held perpendicular to the surface,
we find that
the activation energy barrier for  CO$_{2}$ formation
is reduced to 1.1~eV, and
also that the position of the transition state for the reaction
occurs closer  to the
surface (by 0.3~\AA).
The geometry of the   transition state corresponds to one where 
the optimum tilt angle of the CO axis with respect to the
surface normal is  $49^{\rm o}$ which
represents a ``bond angle''
of  $131^{\circ}$ for the ``CO$_{2}$-like'' complex.
It is interesting to note that this geometry is rather
similar to that associated
with the CO$_{2}^{-}$ ion \cite{bagus} and to that recently proposed
for the ``activated complex'' (or transition state)
 for the CO oxidation reaction over
other transition metal surfaces \cite{coulston}.

In order to understand the origin of the activation energy barrier
for CO$_{2}$ formation,
we analyse the layer-projected density of states (PDOS) 
and charge density and density-difference distributions.
At first it is instructive to consider the PDOS for the free CO and CO$_{2}$
molecules.
Figure~2 shows the result
for CO, where the upper and middle
panels correspond to the
projection of the wave function
of the system onto the atomic orbitals of the
O and C atoms, respectively. In the lower panel we show
the PDOS for projection onto the O adatom.
Similarly, Fig.~3 contains the result for free CO$_{2}$, where
we show the result for just one O atom (the other is identical).
The orbital energy levels of the molecules
are indicated by standard notation appropriate to heteronuclear
systems. 
We note that the energy zero for the PDOS 
of the free molecules have been
aligned with that of the O/Ru\,(0001) surface.
(This is useful for later discussion.)
At this point it is informative to 
briefly describe the PDOS for the O adatom in the
$(1 \times 1)$ O adlayer (lower panel of Fig.~2).
A significant interaction of O 2$p$ and Ru $d$ states
occurs, giving rise to 
bonding and anti-bonding states. These can be seen at energies
of about $-$6.0~eV and 1.8~eV, respectively. In addition we note
the existence of  ``non-bonding'' O 2$p$ states, largely of
$p_{z}$-like character, in the energy region around $-$3~eV.  We have
found that the Ru $d$ states involved
in the bonding with the O 2$p$ orbitals, the linear
combination of which, result in an ``effective'' surface-related state that is
located on the nearest-neighbor Ru atom and has two lobes, one of which
is directed towards the O adatom (the spatial distribution is very similar to that
visible in the right panels of Figs. 5 and 7). 
Furthermore, splitting of the O 2$p_{x,y}$ states can be observed 
as  the peaks in the region $-$6.0 to $-$7.0~eV; 
this is due to the O-O interaction as indicated
by our calculations for a single $(1 \times 1)$-O
layer in vacuum.
We refer to a forthcoming paper in which the
bonding of O on Ru\,(0001) is described for the various ordered
phases that form \cite{forth}.

Considering now the PDOS for the collinear CO transition-state geometry,
as shown in
Fig.~4, from
comparison with Fig.~2 the affect on
the states of the impinging CO molecule and the O adatom
can be observed due to their close proximity.
Significant modifications can be noted:
Firstly, some interaction of CO
with the O adatom 2$s$ orbital can be seen for projection on the C atom
as indicated by the small peak at about $-$19.0~eV.
The CO 4$\sigma$ level has moved to a lower energy 
(by 1.5~eV), as has the 1$\pi$ level (by 1.3~eV). In this respect,
the CO 4$\sigma$ level
appears to be interacting with the O 2$p_{z}$ orbital of the O adatom
as indicated by the small peak at about $-$10.0~eV.  The CO
1$\pi$ state similarly interacts with the O 2$p_{x,y}$ states as
suggested by the small peak at about $-$7.5~eV for projection on
the O adatom.
The most dramatic effect, however, is the 
strong interaction of the CO
5$\sigma$ level  (largely C-like) with the O 2$p_{z}$ states which gives 
rise to new bonding and anti-bonding levels at
about $-$8.0~eV and 3.0~eV, respectively.
(It is to be noted that part of the CO 3$\sigma$-related
 state in Fig.~4 has been cut-off at high binding energy due to the too
 small energy window taken.)
The various changes described in the PDOS are reflected 
in the density difference distributions. Figure~5 shows the total electron
density and the density difference
in the left and right panels, respectively.
The density difference is obtained as the density of the complete
system minus that
of $(1 \times 1)$-O/Ru\,(0001) and minus that of a free CO
molecule. 
It can be noticed that there is charge depletion from the CO 5$\sigma$ orbitals
and from the O 2$p_{z}$ orbitals of the adsorbed oxygen atom.
An increase in electron density can be seen
in the CO $\pi$ orbitals and in the O 2$p_{x,y}$ components.
We can also see that charge has been transferred into 
 $d$ states of the nearest-neighbor Ru atom. These $d$ states appear to be
the same as those from which charge had
been transfered in the formation of the  $(1 \times 1)$ oxygen structure as
briefly noted above.
This movement of electron charge back towards the
Ru atom weakens the O-Ru bond strength.
These results suggest that the activation energy barrier
is due to Pauli repulsion, predominantly 
between the O 2$p_{z}$ states of the O adatom and the
CO 5$\sigma$ orbital which causes electron transfer 
into the CO $\pi$ orbitals and into the O
2$p_{x,y}$ states as well as back into
neighboring Ru atoms.
Furthermore, we note that the close presence of the closed-shell
CO molecule to the $(1 \times 1)$-O/Ru\,(0001) surface induces an average work function
{\em decrease} of 0.35~eV.
 This is in marked contrast to the $(1 \times 1)$ O adlayer
on Ru\,(0001), and to the other lower coverage
ordered phases, for which the expected {\em increase}  in work
function occurs due to 
the  high electronegativity  of oxygen \cite{madey,surnev,stampfl1}.

Understanding of the origin of the reduction in the activation barrier
for CO$_{2}$ formation (from 1.6~eV to 1.1~eV) on relaxing the 
constraint on the molecular axis and the reason for the tilt
can be obtained by considering Figs.~6 and 7. These figures 
show the associated PDOS and electron density and density difference
distributions, respectively, for the identified ``bent'' transition state.
From Fig.~6 it can be seen that 
the most noticeable difference as compared to the results for the collinear
transition state, is the 
development of states  in the region of the Fermi level (both
above and below). 
These states can be 
seen most clearly for projection on the C and O 2$p_{x}$  orbitals. 
We will come back to this point later.
Futhermore, splitting
of the PDOS of the C and O 2$p_{x,y}$ states due to the reduced symmetry can
be observed.
Similarly to the results of the PDOS for the collinear transition state,
 the CO 4$\sigma$
interacts with the O 2$p_{z}$ state; here however, the interaction is
considerably stronger.
Again the CO 5$\sigma$ interacts with the highest lying O 2$p_{z}$ state
giving rise to bonding and
anti-bonding states. 
On investigating the spatial 
distribution 
of these states in the region of the Fermi level, 
we find that they involve the mentioned Ru $d$ states,
states of the O adatom, and the 2$\pi$ orbitals of CO, where for the latter, only
the ``upper'' (furthest away from the surface) lobe is involved.
Partial electron 
occupation of this previously unoccupied state
gives rise to the ``tilted'' or ``bent'' atomic configuration of the
transition state.
Further understanding of 
these bonding rearrangements can be obtained from the density difference
distributions as given in Fig.~7. The 
redistribution of electron charge is 
similar to that of Fig.~5,
with the notable difference however that 
there is a significantly larger increase of electron density 
into the $\pi$ orbitals of the CO molecule; for the C atom this is for
mainly just  {\em one} of the lobes.
This region of charge increase corresponds to that of 
the states described above in 
the PDOS of Fig.~6 which
form around the Fermi level, i.e. 
those involving the CO 2$\pi$ orbitals.

From the above analysis, it thus appears that the microscopic
picture 
of CO$_{2}$ formation via the scattering of CO is as follows:
There is an initial repulsive interaction of CO with the O-covered
surface due to Pauli repulsion between the closed-shell CO and the 
partially negatively charged O adatom, in particular, 
between
the CO 5$\sigma$ and the O 2$p_{z}$ states. 
This interaction serves to reduce the O-Ru
bond strength.
The repulsive  interaction is noticeably
reduced by a ``bent'' configuration since in this geometry charge transfer
may occur into the unoccupied
2$\pi$ orbitals, thus reducing the repulsion.
Bond formation between CO and O to produce CO$_{2}$ 
 then proceeds by charge transfer
{\em back} from the 2$\pi$ ortital into the region of the new bond.
It is conceivable that this  process, i.e., the
role of the CO 2$\pi$ orbitals and the weakening of the adsorbate-substrate
bonds by charge transfer back towards the substrate due to the
repulsive interaction, may be quite 
general and we speculate that it could in fact be the microscopic process
by which oxidation of CO by adsorbed O usually proceeds also in the case
of the L-H mechanism on transition  metal surfaces.

We now ask whether the scattering reaction mechanism for CO$_{2}$
production,  described
above can explain the unusually high rate reported.
From a  rough estimate of the rate \cite{comment}
we find that it is significantly lower than
that observed in the elevated pressure experiments.
This indicates that this mechanism alone is not responsible for 
the high turn-over frequency of CO$_{2}$.
Nevertheless,
were molecular beam experiments to be performed, CO$_{2}$ formation
via this E-R proccess may be identified.
To understand the reported high rate
we turn to another consideration:
Once a CO$_{2}$ molecule has formed and has left the surface
(e.g. by the above described E-R reaction)
there will be an oxygen vacancy in the adlayer.
This site will be occupied by either CO or $\frac{1}{2}$O$_{2}$
from the gas phase.
Our calculations for the adsorption of the O$_{2}$ dimer 
at such a vacancy showed that it is unstable.
On the other hand,
we find that CO can weakly adsorb in an O vacancy with an adsorption
energy is 0.85~eV. This value is significantly less than that which
we calculate for CO on the clean
surface, and on the surface with adsorbed oxygen, for oxygen coverages
less than half a monolayer \cite{stampfl2}. [On the clean surface, using
the same $(2 \times 2)$ periodicity as the rest of the calculations
(rather than the $(\sqrt{3} \times \sqrt{3})R30^{\circ}$ structure that actually
forms), the value obtained is 1.68~eV; 
the experimental value has been reported to be 1.66~eV~\cite{coexp}.]
Adsorption of $\frac{1}{2}$O$_{2}$ (i.e., an O atom from O$_{2}$)
in such a vacancy is more favorable, where the
adsorption energy is 1.20~eV. Assuming thermal equilibrium and
using the law of mass action \cite{commentmass} the calculated
energies imply that 0.03~\% of sites in the O adlayer will be occupied
by CO.
Chemisorption of $\frac{1}{2}$O$_{2}$ in the vacancy, however,
from O$_2$ gas will be hindered by an energy 
barrier for dissociation~\cite{over,weinberg} whereas CO may occupy this
site essentially without hindrance. 
This means that the actual concentration of adsorbed CO will be somewhat higher
than that estimated above.
Reaction to CO$_{2}$ from CO at the vacancy and a neighboring
O adatom (the surface reaction step of the heterogeneous cycle),
is calculated to give rise to an energy gain of
$\approx$0.66~eV.
This 
value is obtained as the difference between 
the reaction energy in vacuum and the sum of the adsorption energies
of O and CO on the surface prior to reaction.
This value is noticeably smaller than that of 1.95~eV  as we obtained 
for the E-R mechanism,
but is still quite large if compared to that of
 $\approx 0.2$~eV as obtained experimentally 
for CO oxidation over Pt\,(111) and Pd\,(111) \cite{king}.
We do note that some of this energy may be disipated to phonons, thus
reducing the value somewhat.
A L-H reaction between the adsorbed CO molecule in a vacancy
and a neighboring O adatom is expected to be
particularly efficient and to give rise to the high rate
 due to the mentioned weaker bond strengths of
{\em both}
 the adsorbed CO molecule and the O adatom, as well as their  close proximity 
to one another.
As noted in the introduction, under UHV conditions, ruthenium
is the least efficient
catalyst for the CO oxidation reaction. This is likely to be related to
the fact that for low oxygen coverages
 ($\Theta \leq 0.5$) as occurs under
UHV,  ruthenium binds oxygen (and carbon monoxide) particularly strongly.
From simple arguments, 
this is likely to lead to a lower reaction rate as compared to that over
other transition metal catalyts, which under the same condidtions,
bind these particles less strongly.


\begin{figure}
\psfig{figure=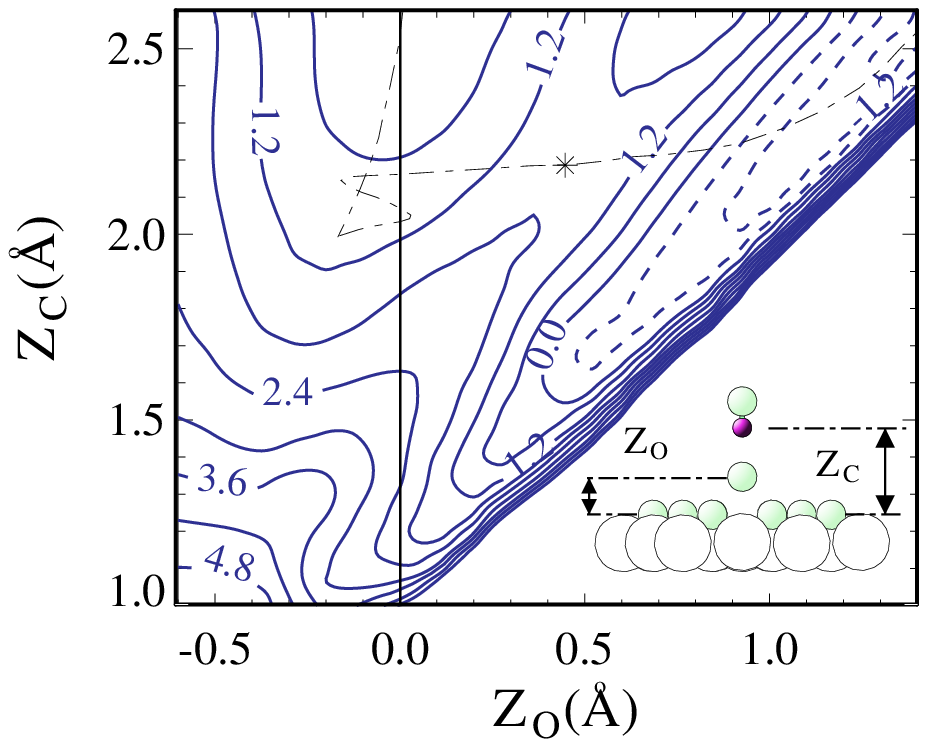}
\caption{Potential energy surface (PES) 
as a function of the positions of the C atom, $Z_{\rm C}$, 
and of the O adatom, $Z_{\rm O}$.
Positive energies (i.e. repulsion) are shown as continuous lines,
negative ones  (i.e. energy gain) as
dashed lines.
The contour-line spacing is 0.6 eV.  }
\end{figure}

\clearpage

\begin{figure}
\psfig{figure=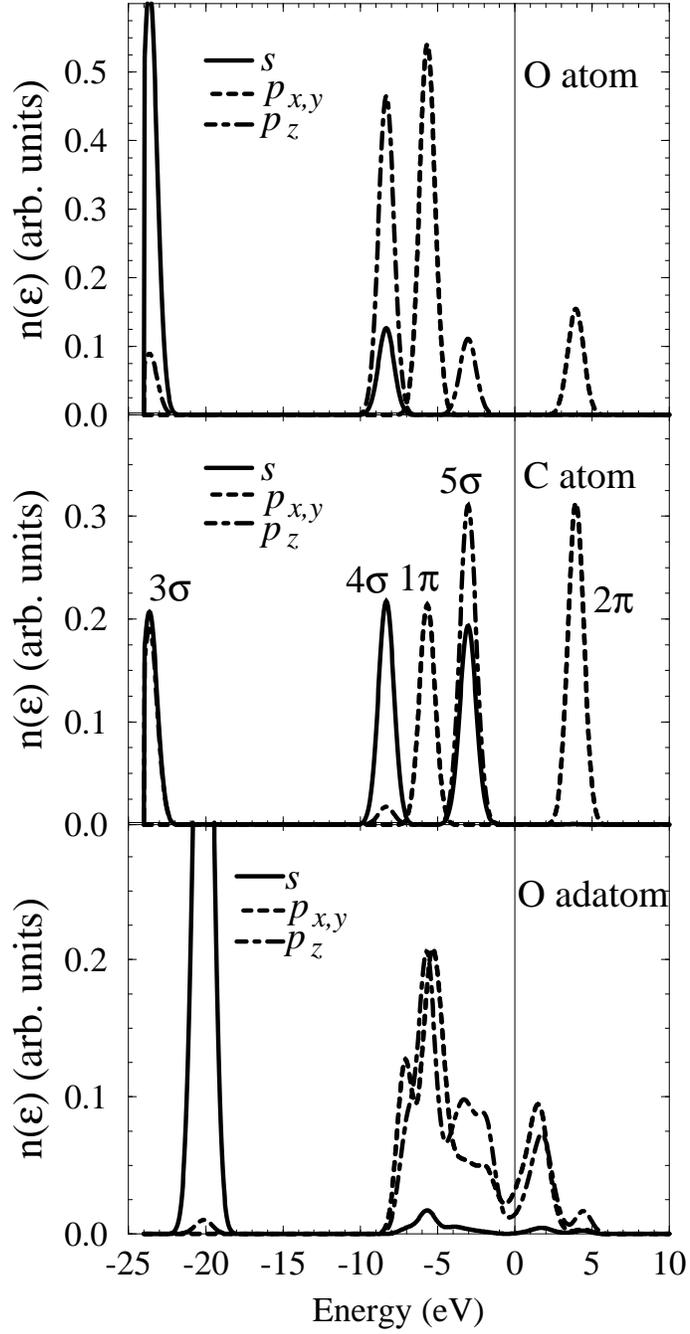}
\caption{Projected density of states (PDOS) for a free  CO molecule.
The upper,
middle, and lower panels correspond to projection on the O and C atoms of CO, 
and on the O adatom, respectively.}
\end{figure}

\clearpage

\begin{figure}
\psfig{figure=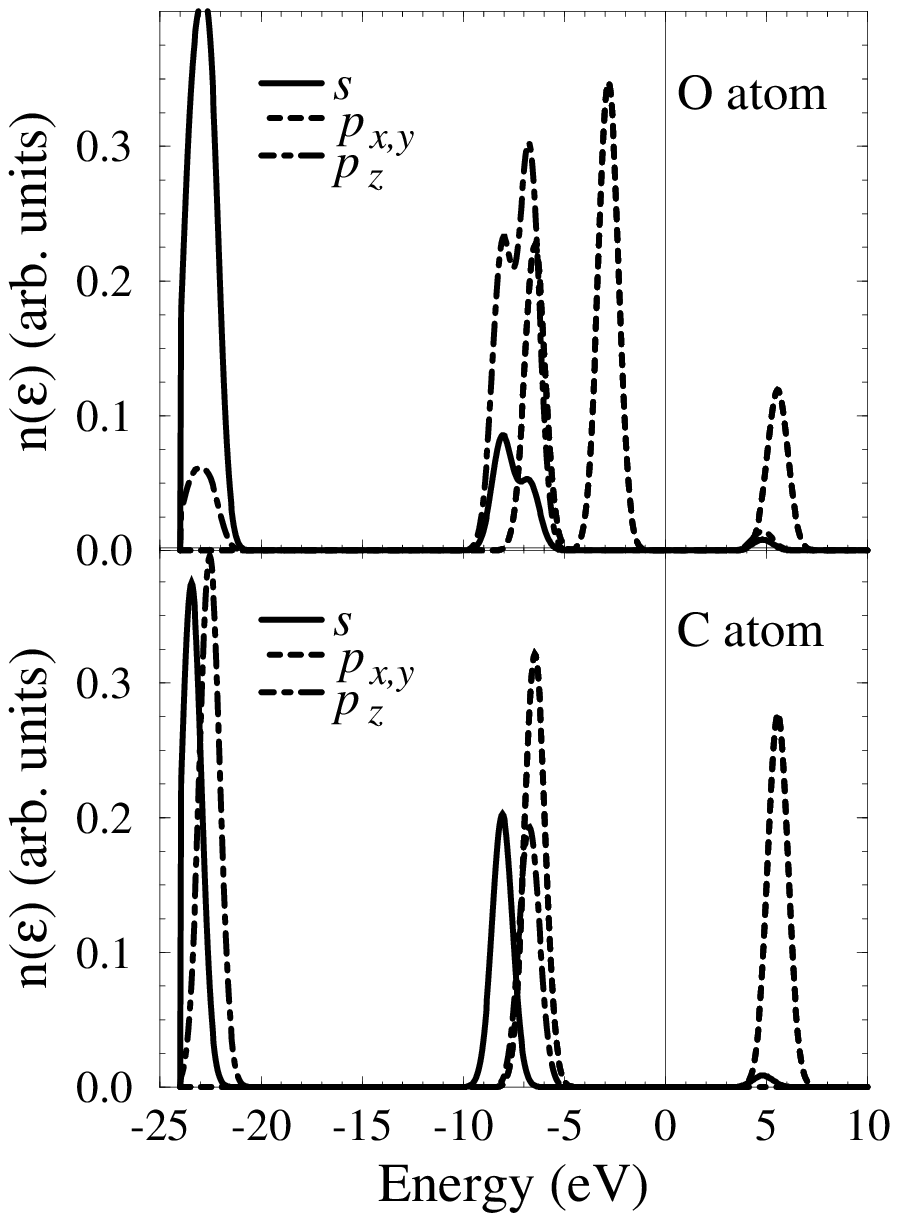}
\caption{Projected density of states (PDOS) for a free CO$_{2}$ molecule.
The upper and lower panels correspond to projection on the O and
 C atoms, respectively.}
\end{figure}

\clearpage

\begin{figure}
\psfig{figure=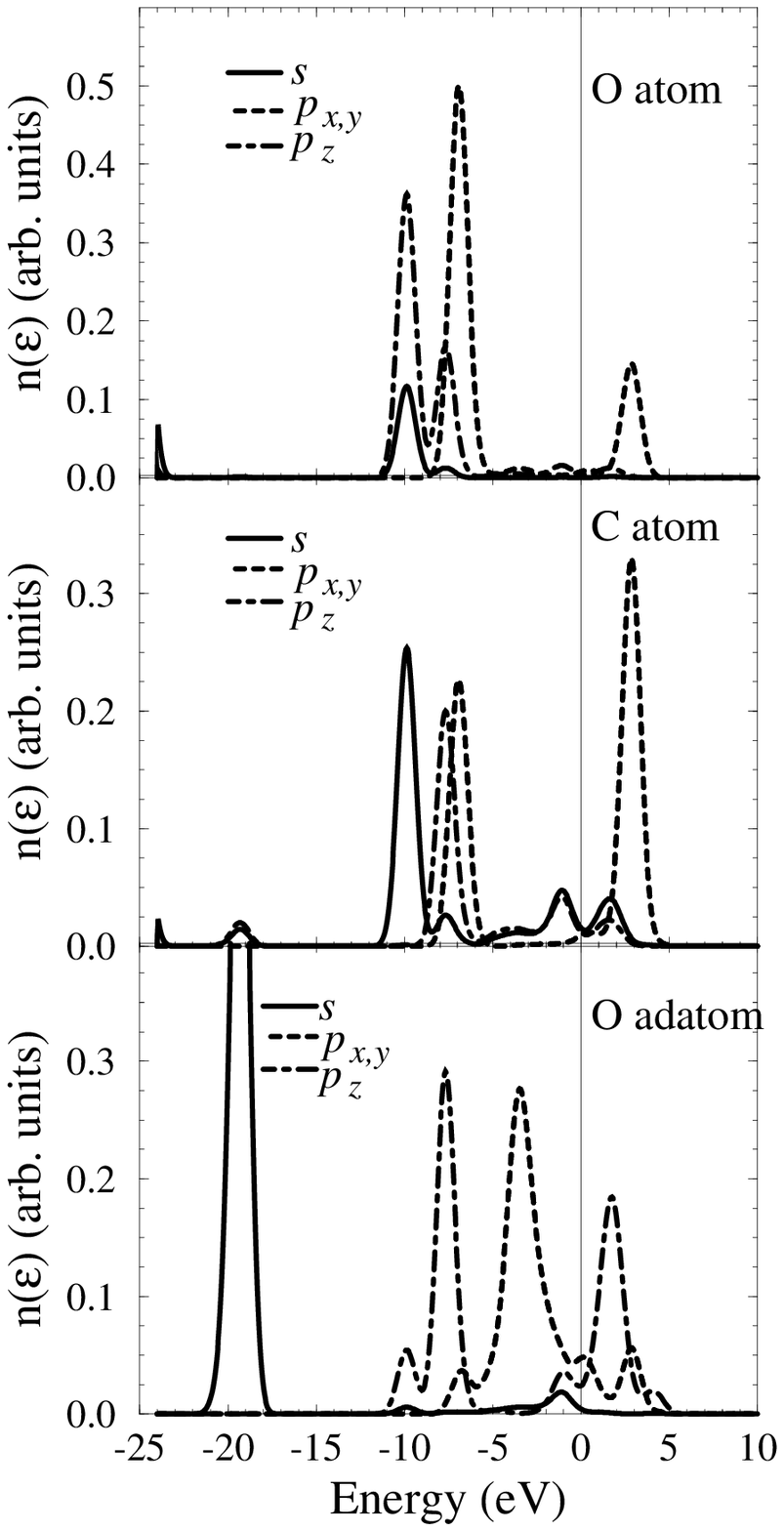}
\caption{Projected density of states (PDOS)
 for the ``collinear'' transition state for 
reaction via scattering of CO from $(1 \times 1)$-O/Ru\,(0001). The upper,
middle, and lower panels correspond to projection on the O and C atoms of CO, 
and on the O adatom, respectively.}
\end{figure}

\begin{figure}
\psfig{figure=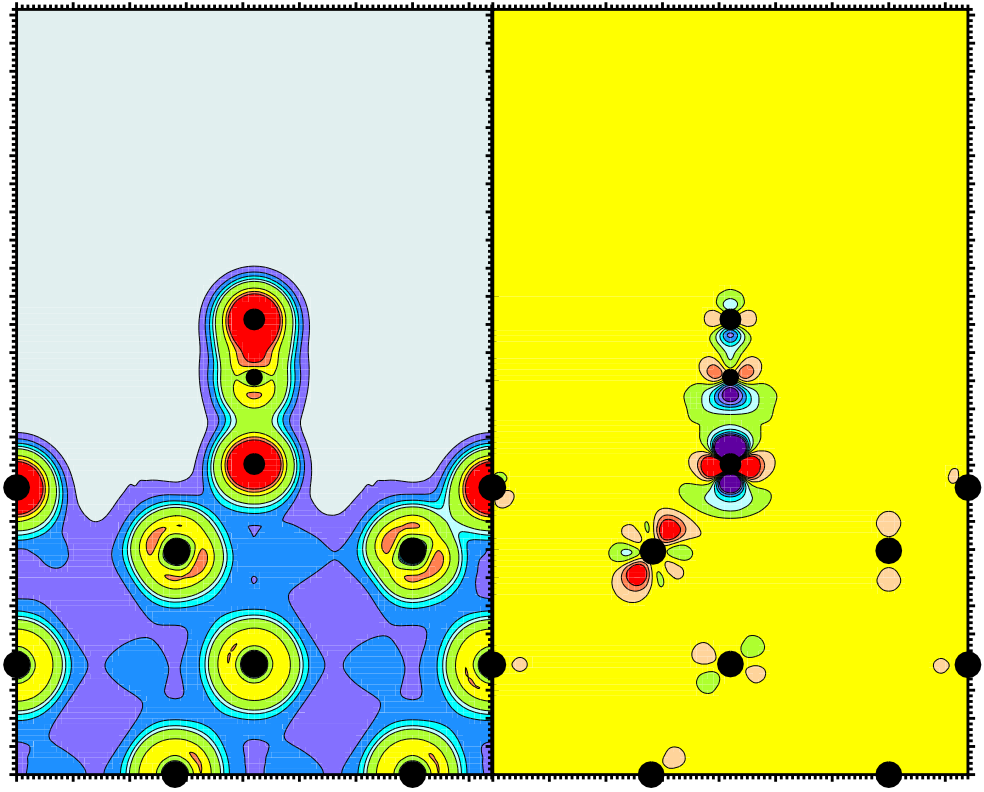}
\caption{Electron density (right panel) and density difference
(left panel)  distributions for the ``collinear'' transition state.
The contours for the electron density are
such that the spacing of the first five contour lines is
2.0, where the first line begins at 2.0. Thereafter the spacing is 10.0.
For the density difference, the contour spacing is 0.6 where the first positive countour line begins at 0.3 and the first negative one at $-$0.3. 
The warm (orange, red) and cool (green, blue, purple) colors
represent regions of charge increase and decrease, respectively.
The units are $10^{-2}e/a_{0}^{3}$.}
\end{figure}

\clearpage

\begin{figure}
\psfig{figure=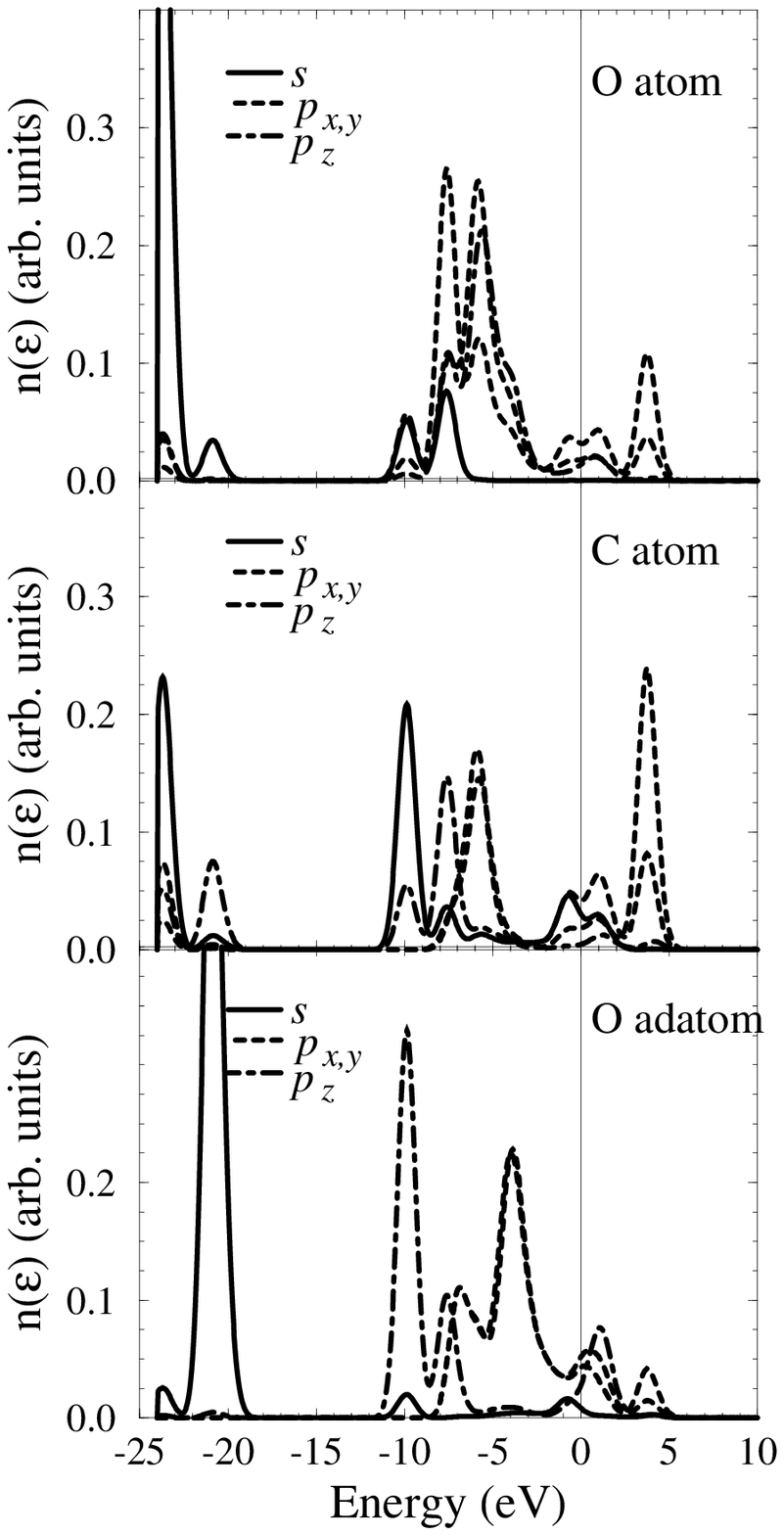}
\caption{Projected density of states (PDOS)
 for the identified ``bent'' transition state for 
reaction via scattering of CO from $(1 \times 1)$-O/Ru\,(0001).
The upper,
middle, and lower panels correspond to projection on the O and C atoms of CO, 
and on the O adatom, respectively.}
\end{figure}

\clearpage

\begin{figure}
\psfig{figure=fig7.ps}
\caption{Electron density (right panel)  and density difference 
(left panel) distributions
for the identified ``bent'' transition state. The contour lines are
the same as given in Fig.~5.}
\end{figure}


\begin{references}
\bibitem{comment} For
our estimate we use the frequency of impinging CO molecules  per surface
Ru atom of
$\nu = 7.5 \times 10^{6}\, s^{-1}$
as the prefactor which corresponds to a CO gas phase pressure of
16 torr, and the temperature is taken to be  $T = 500~$~K.
\bibitem{commentmass}
We assumed that the
O$_2$ and CO partial pressures are equal, the temperature
is $T = 500$~K,
and the binding energy of CO into an O-vacancy in  the
$(1 \times 1)$ O adlayer is calculated to be by 0.35 eV less favorable than
that of $\frac{1}{2}$ O$_2$.
\end{references}
\end{document}